# Relaxation models for Single Helical Reversed Field Pinch plasmas

R. Paccagnella^

Consorzio RFX, Corso Stati Uniti 4, 35127 Padova, Italy

^and Istituto Gas Ionizzati del CNR

**Abstract:**

In this paper a relaxation theory for plasmas where a single dominant mode is present [1], is revisited. The solutions of a related eigenvalue problem are numerically calculated and discussed. Although these solutions can reproduce well the magnetic fields measured in experiments, there is no way within the theory to determine the dominant mode, whose pitch is a free parameter in the model. To find the preferred helical perturbation a procedure is proposed that minimizes the "distance" of the relaxed state from a state which is constructed as a two region generalization of the Taylor's relaxation model [2,3] and that allows current discontinuities. It is found that this comparison is able to predict the observed scaling with the aspect ratio and reversal parameter for the dominant mode in the Single Helical states. The aspect ratio scaling alone is discussed in a previous paper [4] in terms of the efficient response of a toroidal shell to specific modes (leaving a sign undetermined), showing that the ideal wall boundary condition, a key ingredient in relaxation theories, is particularly well matched for them. Therefore the present paper together with [4] can give a new and satisfactory explanation of some robust and reproducible experimental facts observed in the Single Helical (SH) Reversed Field Pinch (RFP) plasmas and never explained before.

**Introduction:**

In a previous paper [4] (hereafter called paper I) the numerical and experimental discovery of the Single Helical states in Reversed Field Pinches (see Ref.s [1-8] therein) is discussed in some detail.

In paper I, the failure of Taylor's relaxation theory [2,3] in predicting the observed dominant helicities is also analyzed and it is shown that the main obstacle is due to the fact that Taylor's theory predicts a flat parallel current density in the core of a force-free plasma, while the observed modes are triggered either by a current or by a pressure gradient localized near to the magnetic axis. In the same paper it is further shown that a toroidal conducting shell is able to respond efficiently to the modes observed in RFPs in SH states at shallow reversal (i.e. having a toroidal field reversal radius very close to the wall). For these modes the wall is inductively optimally coupled and passively counter reacts the field penetration as a quasi- ideal wall (producing a zero or very low radial magnetic field at the wall radius).

Since the presence of an ideal shell surrounding the plasma is crucial to preserve global invariants (like total helicity, magnetic energy, magnetic flux etc.) these findings suggest to consider some relaxation mechanisms, although different from the Taylor's one, responsible for the excitation of the experimentally observed helicities.

A few years ago a theory was proposed which appears appropriate to describe situations in which a single mode becomes dominant [1]. Unfortunately at that time the SH helical states (see paper I for references) were not yet been observed. Moreover Ref. [1] proposed a set of equations which are not easily solved as an eigenvalue problem being quite sensitive to the initial condition (at least in some parameters regions). Also, within this theory the helical pitch of the dominant mode is a completely free input parameter.

In more recent years some papers discussed the relaxation process in presence of current sheets [5, 6] or in multiple regions [7]. Clearly in resistive plasmas during SH states island like structures could develop, and, as a consequence, the idea of multiple relaxation regions becomes natural. Resistive magneto-hydro-dynamic (MHD) modes, that can be associated with islands, can act as a barrier for a complete relaxation of the system toward a Taylor's state.

In this paper we discuss in some detail the solutions of the eigenvalue system proposed in [1]. Afterwards we construct a two region generalization of the Taylor's model, where, one region is delimited by the current island location in the plasma core and the second extends from the island edge to the wall. By defining an appropriate measure of the "distance" between the magnetic field profiles predicted by the two models we show that the minimum distance can determine the toroidal mode number of the favorite dominant mode.

The paper is organized as follows: in section I the helical relaxation model [1] is revisited; in section II a two region relaxation model is formalized; in section III the procedure to determine the dominant helicity as a function of the aspect ratio and reversal parameter is presented and finally general discussion and conclusions are given.

## I. The Single helicity relaxation model

In this section the model described in [1] and the procedure to obtain a numerical solution of the associate eigenvalue problem is briefly summarized.

The relaxation model assumes a dominant helicity in the plasma and a minimization procedure for energy subjected to the conservation of total helicity (as in Taylor's theory [2]) and of an invariant related to the dominant mode expressed as [8]:

$$K_1 = \frac{1}{2}\int_V \chi^d \mathbf{A} \cdot \mathbf{B}\, dV \qquad (1)$$

where the integral is over the whole plasma volume, **A** and **B** are respectively the magnetic potential and the magnetic field and $\chi$ is the helical flux function of the mode defined as:

$$\chi = q_s \Psi - \Phi \qquad (2)$$

where $q_s$ is the mode pitch , $\Psi$ is the poloidal flux, while $\Phi$ is the toroidal flux, d is a positive integer. The new invariant in Eq. (1) corresponds therefore to the total helicity "weighted" over some power of the helical flux (and hence of the helicity) of the dominant mode. The underlying idea is that together with the total helicity, the invariant $K_1$ is also a well preserved quantity in a relaxation process dominated by a single mode. Note that the exponent d was not considered in the 1980 paper [1]. However later in Ref.[8], the authors realized that this parameter was needed in order to obtain solutions independent of the normalization of the magnetic field. This is clearly a very important and necessary physical constraint. Therefore although the degrees of freedom of the theory increased by adding an extra unknown parameter , there was no possible alternative. As it will be discussed later, this new free parameter will play an important role to improve the comparison of the relaxed state solutions with the experimental data.

The relaxed state (assuming a vanishing current at the plasma edge) is shown [8] to satisfy the equation:

$$\boldsymbol{J} = \frac{\lambda_0 \ (d+2)}{2} \chi^d \boldsymbol{B} \qquad (3)$$

where $\lambda_0$ is a constant. As discussed in [1] Eq.(3) can be solved in cylindrical geometry as a two point boundary value problem for an ODE system of four coupled variables with conditions at r=0 and at r=1 (by normalizing the radius r to the plasma minor radius a). It should be mentioned that the numerical solution of this system is not trivial, since it shows a behavior, at least in some intervals of the free parameter $\lambda_0$ , quite sensitive to the initial condition. This sensitivity shows up, in particular, in the fact that a good convergence to the final eigenstate can be obtained numerically only if the initial state is not too " faraway" from the final solution; similarly to what can happen sometime in searching the roots of a nonlinear algebraic equation where the initial guess could be critical for the final convergence.

As discussed in [1,9] there are pinch like (P) and tokamak like (T) solutions of the system. The solution only depends on $q_s$. However since $q_s$ = (R/a ) m/n where R/a is the torus aspect ratio while m and n are respectively the poloidal and toroidal mode numbers, the aspect ratio and mode helicity can be chosen separately.

I the next section several interesting features of the solutions described by Eq.(3) (by looking in particular to the P branch) are described.

### I.1 Single helical relaxed eigenstates properties

The solutions of Eq.(3) (or of the associated fourth order ODE [9] ) for the P branch are very rich.

A full description and analysis of them is however not the main purpose of this paper,

therefore here, only a few results, that are relevant for the following discussion, are considered.

Assuming, as in [9], d=1 in Eq.(3), an example of the relaxed profiles is given in Fig.1, where four cases at different values of the reversal parameter F (i.e. the ratio between the toroidal field at the wall and the average toroidal field over the plasma cross section) are shown. As in the rest of the paper, we set m=1, in agreement with the experimental evidence during SH states. In this way the relaxed state can be determined by selecting n, the toroidal mode number, and the aspect ratio.

Note that the negative sign in front of n in Fig.1 selects the modes resonant near the magnetic axis, in the core plasma.

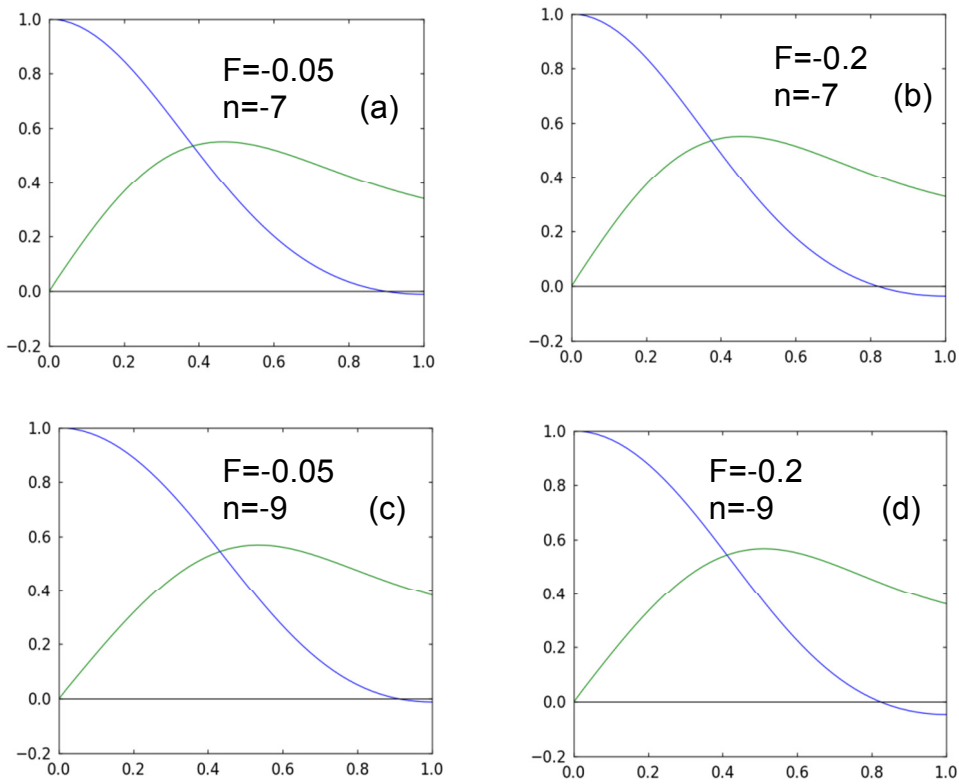

*Fig.1: Poloidal (in green) and toroidal field (in blue) components as a function of normalized radius r/a corresponding to the eigenstates with eigenvalues : $\lambda_o$ = -2.3 (a), -2 (b), -3.9 (c) ,-3.1 (d) for an aspect ratio R/a=5 (assuming m=1). These cases are obtained with d=1 and J(a)=0 as in Ref.[1].*

An important feature for determining the stability of the magnetic field profiles in RFP's, as discussed in detail in paper I, is the parallel current profile plotted in Fig.2 for one of the cases of Fig.1.

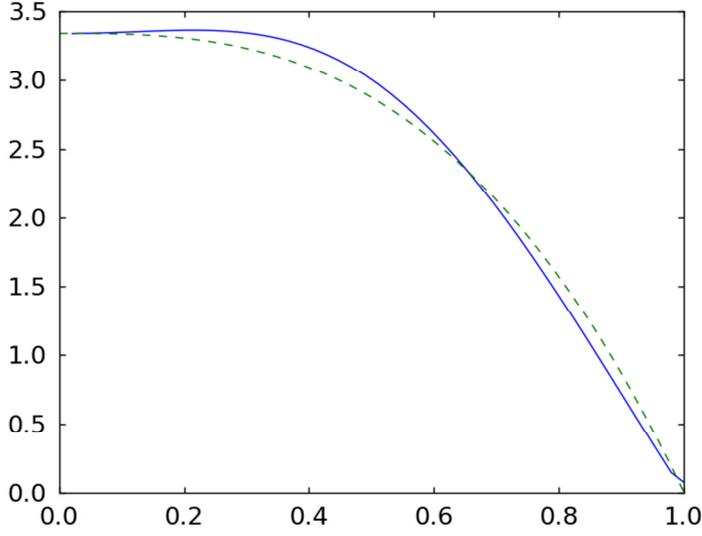

***Fig.2:*** *Parallel current radial profile $\alpha = \mathbf{J} \mathbf{B}/B^2$ for the case n=-9 F=-0.05 of Fig.1. The dashed line is a fit (see text).*

In Fig.2 a fit of the parallel current (dashed line) is also given by employing the function :

$\alpha(r) = \alpha(0) (1-(r/a)^{\alpha_{fit}})$ with $\alpha(0)$ as obtained from the relaxed profiles solution and $\alpha_{fit}$ being a free parameter. In the case of Fig.2 $\alpha_{fit} = 2.83$ is found. In general, for a given aspect ratio, the flatness (corresponding to more stable profiles) increases with n as shown in Fig.3. Note that, as in the rest of the paper, absolute n's values are shown, referring however always to modes resonant in the core (i.e with negative n's values).

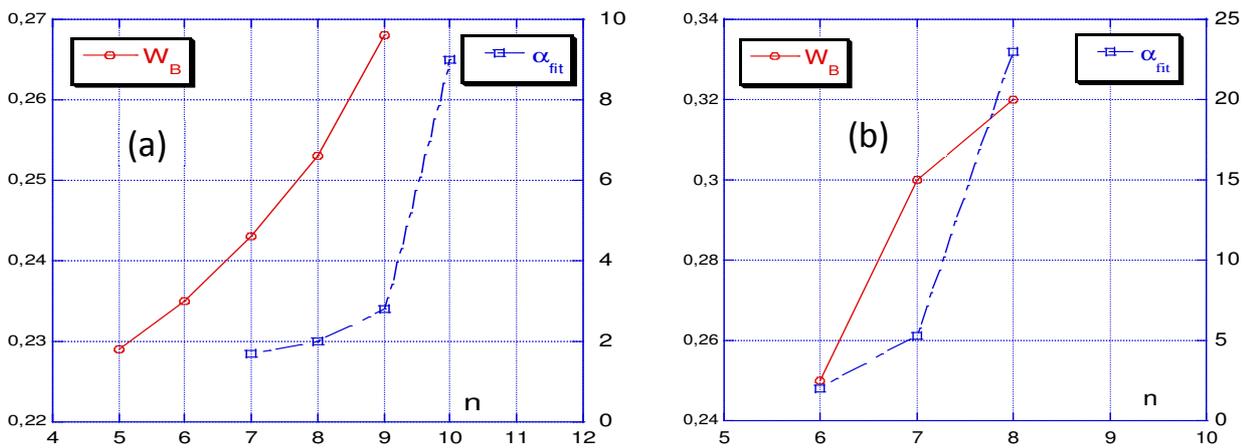

***Fig.3*** *Magnetic energy $W_B$ and fitting exponent of the $\alpha$ profile vs. n for R/a=5 and F=-0.05 (shallow reversed case) (a) with d=1 as in Ref.[1] and (b) with d=2 as in Sec. III (absolute n values are shown).*

Note that for $\alpha_{fit}$ less or around 2 (and $\alpha(a)=0$) the configurations are ideally unstable [10].

It should be remarked that the stability of the axi-symmetric equilibria (as done in [10]) is relevant here, since the relaxed states obtained from Eq.(3) are in fact axi-symmetric.

Therefore assuming d=1 and the same profiles as in Ref.[1] (Fig.3(a)) n=7,8 modes likely correspond to unstable configurations. Instead for d=2 and a finite value of the parallel current at the wall (see also Sec.III) all cases are stable (Fig.3(b)).

In Fig. 3 the total magnetic energy, i.e. the integral over the volume of the squared magnetic field amplitude, obtained with different n's and calculated for the eigenstate solutions, is also shown. It can be seen that it monotonically increases with n. These general features are confirmed for different aspect ratios and F values. In general at low aspect ratio, R/a=2, the increasing rate as a function of n of $W_B$ and $\alpha_{fit}$, is even steeper than for the case shown in Fig.3.

On the basis of this brief analysis of the properties of the solutions of Eq.(3) we can conclude that, assuming d=1 and zero current density at the plasma edge, some of the relaxed states solutions correspond to stable magnetic fields (at relatively high n's) while others (typically at low n's) can be ideally unstable. The magnetic energy, which has obviously a local stationary point subjected to the given constraints for each eigenstate and corresponding eigenvalue, is nevertheless a monotonically increasing function with the toroidal mode number.

Similar properties, corresponding however to fully stable configurations, are obtained with d=2 and with a finite parallel current density near to the wall, a case that will be further exploited in Sec.III.

In Fig.4 the F vs. Θ curves ( F /Θ being the ratios between the toroidal/poloidal magnetic field at the wall to the total toroidal magnetic flux over the plasma volume) are shown for different aspect ratios and n's values. It should be noted that the values of Θ at which F=0 (i.e. when the magnetic toroidal field reverses exactly at the wall) depend on both R/a and n. As already noted in [1], they match much better, the experimental operational range, than the Taylor's fully relaxed states. The case (b) in Fig.4 corresponds to the more realistic interval of F and Θ values (if compared with experiments) and, as mentioned, also to stable profiles over the whole range of n values.

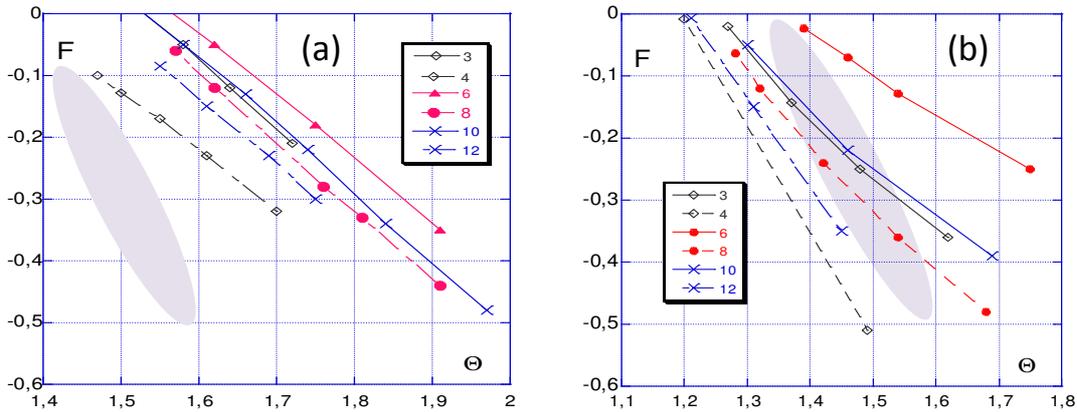

***Fig.4:*** *F vs Θ for different aspect ratios (R/a=2 diamonds, 5 plain circles and 6.8 crosses) and n's (a) with d=1 as in [1] and (b) with d=2 as in Sec.III. The shaded region in (a,b) corresponds approximately to typical RFX-mod data.*

Finally it has been checked that typically the case with d=3 gives unrealistic magnetic field profiles (quite peaked on axis) and no toroidal field reversed (see Fig.5 as an example) cases have been obtained.

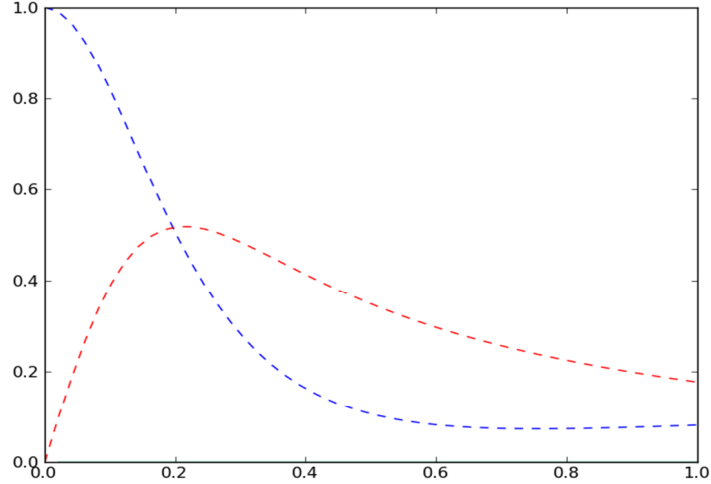

***Fig.5***: *Poloidal (in red) and toroidal field (in blue) components as a function of normalized radius r/a, with d=3 (for a case with R/a=2 n=3)*

## II. A two region Taylor's relaxation model

SH states in RFP's correspond to fully reconnected helical plasmas. The presence of a reconnecting mode and the corresponding island structure in the SH states justifies therefore the idea to consider a generalization of Taylor relaxation in which the internal region (approximately within the X-point of the island) corresponds to a different relaxed state with respect to the external region (from the X-point to the wall). Clearly this is not a rigorous relaxation theory, however it is the most natural and simple generalization of Taylor's force-free states allowing a finite parallel current gradient, which in turn, as discussed in the Introduction (and more in detail in paper I) is essential in order to permit the destabilization of the internal resonant modes, i.e. those actually observed in experiments.

The presence of a reconnecting layer, on the other hand, produces a discontinuity that has to be handled in some way [5,6,7]. Although maybe too simplistic, the model presented here is a possible (and physically reasonable ) way to deal with this important issue avoiding complicate formalisms [5].

We assume that the plasma will satisfy a force-free relaxed state solution:

$$(\nabla \times \boldsymbol{B})_\| = \alpha(r) \boldsymbol{B} \qquad (4)$$

To construct the solution over the whole radius we divide the plasma in a core region, $0 \leq r < r_c$,

where α is exactly constant with a value $\alpha_1$, while for $r_c < r \leq a$, we express the magnetic field as a combination of Bessel Functions with a different constant $\alpha_2$.

With these assumptions the full solution of Eq.(4) in cylindrical geometry and axi-symmetry (all fields are only functions of the radius ( r ) of the cylinder) can be written as:

$$\mathbf{B} = (0, J_1(\alpha_1 r), J_0(\alpha_1 r)) \quad \text{in } 0 \leq r < r_c$$

$$\mathbf{B} = (0, b_1 J_1(\alpha_2 r) + b_2 Y_1(\alpha_2 r), c_1 J_0(\alpha_2 r) + c_2 Y_0(\alpha_2 r)) \quad \text{in } r_c < r \leq a \quad (5)$$

where the J's and Y's are Bessel functions of the first and second kind respectively and b's and c's are arbitrary constants to be determined by matching the solutions in $r=r_c$ imposing continuity of the fields and their derivatives.

Simple algebraic relations can be deduced for the constants:

$$b_1 = \frac{1}{2} \pi r_c \left( \alpha_2 J_1(\alpha_1 r_c) Y_0(\alpha_2 r_c) - (k1 + \alpha_1 J_0(\alpha_1 r_c)) Y_1(\alpha_2 r_c) \right)$$

$$b_2 = \frac{1}{2} \pi r_c \left( -\alpha_2 J_0(\alpha_2 r_c) J_1(\alpha_1 r_c) + (k1 + \alpha_1 J_0(\alpha_1 r_c)) J_1(\alpha_2 r_c) \right)$$

$$c_1 = \frac{1}{2} \pi r_c \left( (-k0 + \alpha_1 J_1(\alpha_1 r_c)) Y_0(\alpha_2 r_c) - \alpha_2 J_0(\alpha_1 r_c) Y_1(\alpha_2 r_c) \right)$$

$$c_2 = \frac{1}{2} \pi r_c \left( (k0 - \alpha_1 J_1(\alpha_1 r_c)) J_0(\alpha_2 r_c) + \alpha_2 J_0(\alpha_1 r_c) J_1(\alpha_2 r_c) \right) \quad (6)$$

These expressions are written for the general case where a local jump in the derivative of the fields is present. The amplitude of the jump is determined by specifying the two free constants k0, k1 related to the discontinuities of the poloidal and toroidal fields derivative respectively. Although the current jump does not play a role in our results, interestingly, it has been shown [11] that by modulating the jump it would be possible to directly affect the plasma stability. Therefore we add this degree of freedom in our model. Moreover the constants k0 and k1 could also be used (see below) to adjust the edge current density.

Examples of the toroidal and poloidal magnetic field profiles obtained with this model are given in Fig.6 for different sets of the free parameters $r_c$, $\alpha_{1,2}$.

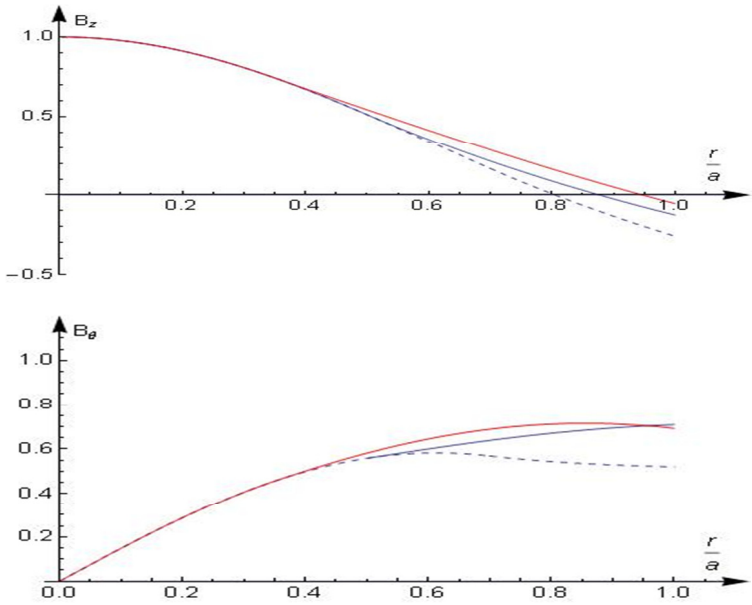

*Fig.6:* Toroidal(top) and poloidal (bottom) magnetic field profiles vs. r/a. From the dashed line toward less reversed cases the parameters are: $r_c$ =0.7, $\alpha_1$=3 , $\alpha_2$=0.5 (dashed curve) ; $r_c$ =0.5, $\alpha_1$=3, $\alpha_2$=1.5 (blue curve) and $r_c$ =0.35, $\alpha_1$=3, $\alpha_2$=2.1 (red curve) and k0=k1=0 (no current jump).

By construction the two parameter model has an $\alpha$ profile which is a function of radius (see Fig.7). The fact that $\alpha$ becomes a decreasing function of radius is, as discussed in paper I, a desirable feature to affect the stability of the on-axis resonant modes which are instead completely stabilized for a constant $\alpha$ profile.

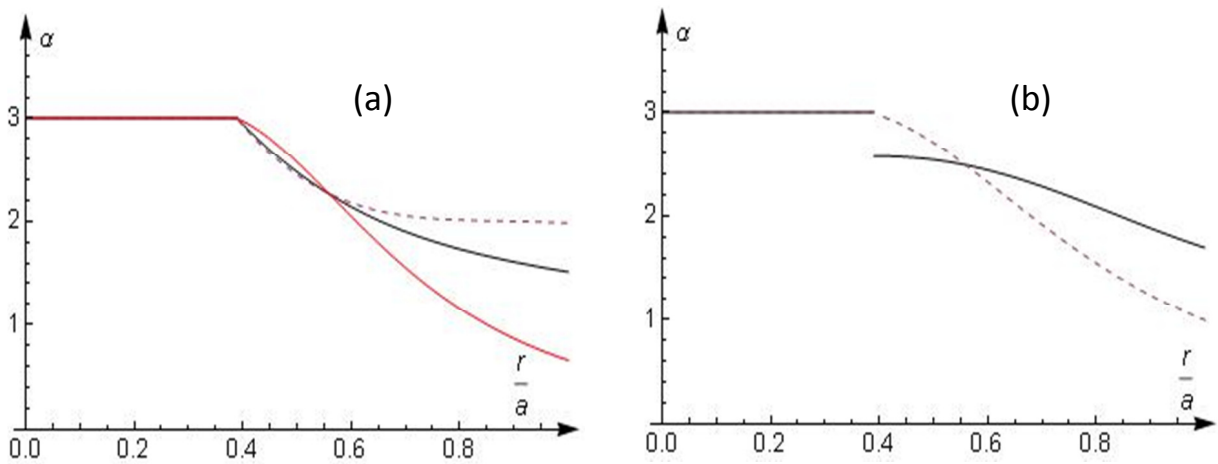

*Fig.7:* $\alpha$ vs. normalized radius for $r_c$ =0.39 with $\alpha_1$ =3 and $\alpha_2$ = 2.4(dashed curve) , 2.1 (black) , 0.1(red) and k0=k1=0 (a) and k0=0.2,k1=-0.3 ((b) plain line) and k0=0.3,k1=0.2 ((b) dashed line). Note that the discontinuity in (b) is introduced only for illustration purposes.

One important thing to note in Fig.7(a) is that the variation of the parallel current profile in the external region (i.e. for r > $r_c$) is not extremely influenced by the choice of the $\alpha_2$ parameter. For $0 < \alpha_2 < 0.1$ the profile does not change (it saturates toward the red curve of the figure), while for $\alpha_2 > 2.4$ it becomes non monotonic (not shown).

The jump in Fig.7(b) is introduced for illustration purposes, however, the jumps in the derivatives at $r=r_c$ can be used either to produce a real jump in the parallel current profile (that can determine plasma stability [11]) or to modulate the parallel current gradient in the region with $r > r_c$ and its value at the wall. Due to the presence of reconnecting resistive modes it should be expected that current sheets (i.e magnetic field discontinuities) could develop. Moreover due to differences in the transport mechanisms inside and outside islands also current jumps could possibly arise in real cases. Therefore the two region relaxation model has been kept general enough to be able to describe all these cases.

### III. Dominant mode scaling with R/a and F

In this section a procedure is established in order to determine the toroidal mode number, n, that emerges at different aspect ratios and also at different F values.

As described in Sec.I, within the single helical mode relaxation theory [1,9], the n (assuming m=1) value is just an input parameter. It has also shown that for some aspect ratios and n values (with d=1) the profiles are unstable, since they correspond to relatively peaked parallel current profiles. Within this theory only a smooth relaxation process without current sheets is described.

On the other hand, to describe situations where reconnecting modes and island structures develop we have constructed a flexible model that generalizes Taylor's states to the case where two separate relaxations regions exist: a fully relaxed state with flat current within the island and a partially relaxed state elsewhere.

In the following it is shown that it is possible to determine the favorite/dominant n mode, by mixing the two approaches i.e. assuming that the emerging states, at each aspect ratio and F values, are those for which the distance (in some sense discussed later) between the profiles that can be obtained by the relaxation theory [1,9] and those that result from a partial relaxation that allows the presence of islands (and/or current discontinuities), as described in Sec.II, is minimum.

Finite currents jumps are allowed in the model, however, it will be shown that it is not necessary to specify the amplitude of the jump, since interesting results can be obtained even in the limit k0=k1=0 (see Sec. II).

Note that, in the following, the results of the single helical relaxation model are obtained setting d=2 in Eq.(3) and a finite parallel current at the edge. As shown in Sec.I these profiles are stable for every n values and are also covering a region in the F-Θ operational space well overlapped to that of the existing experiments. The profiles deduced by solving Eq.(3) and those obtained from the two region relaxation model are similar, as can be seen in Fig.8.

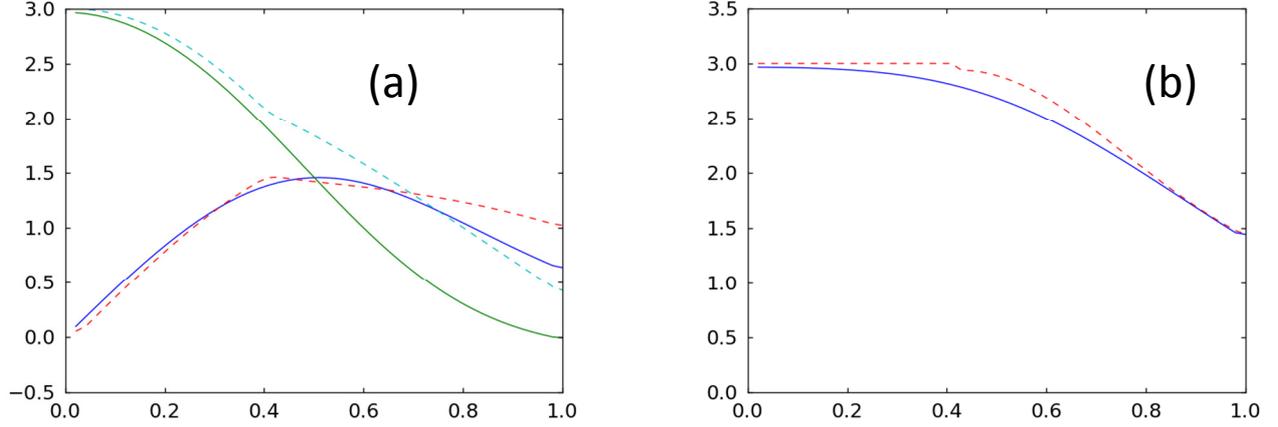

***Fig.8:*** *Poloidal and toroidal current densities (a) and $\alpha$ profile (b) vs. normalized radius obtained from Eq.(3) (plain lines) with R/a=5,n=-7, F=-0.05 and d=2 and from the two region model (dashed lines) with $\alpha_1$ =3 , $\alpha_2$ =2.1 and $r_c$ =0.39.*

The differences between the two models are appreciable mainly in the external region due to a different behaviour of the current profiles components. Note however that the parallel current $\alpha$ profile (Fig.8b) are very similar. Therefore, for example, the stability properties should be expected to be practically identical.

An interesting feature, is however, that the "mean" difference between the two region and the helical relaxation models, depends on the helical mode number. In fact by defining the integral error function:

$$E(q_s) = \int (J_\|^{\alpha_{1,2}} - J_\|^{\chi})^2 \, dr$$

where $J_\|^{\alpha_{1,2}}$ is the parallel current deduced from our 2 region model, and $J_\|^{\chi}$ is the parallel current has deduced from Eq.(3) and the integral is over the whole normalized radial extent (from 0 to 1); the aim is, as mentioned, to find a value of the pitch, $q_s$, which minimizes this difference. In this way for each aspect ratio (and also F value) the helical solution nearest to our two region relaxation model can be found. This procedure ultimately overcomes the limitation of the helical relaxation model [1,9] by determining the helicity and allowing, at the same time, the presence of currrent jumps. The numerical results are shown in Fig.9.

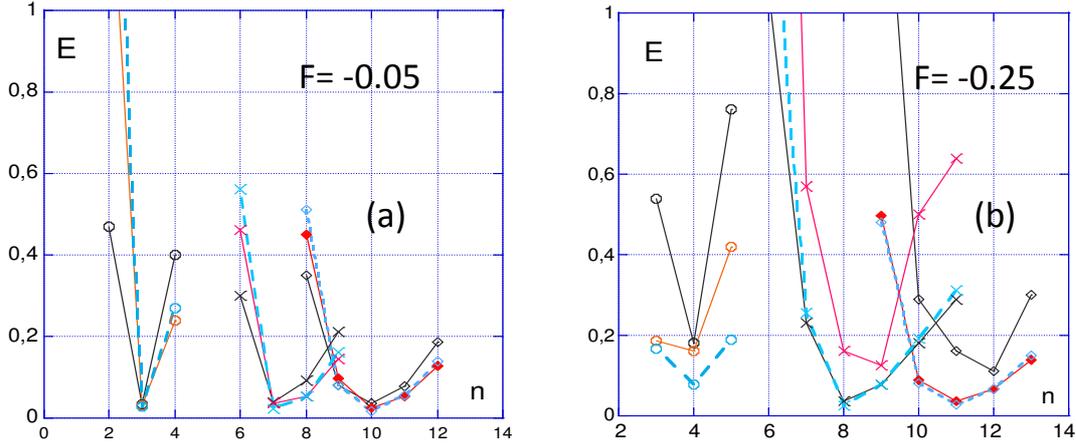

***Fig.9:*** *The integral difference E vs. n for different aspect ratios (R/a=2 circles, 5 crosses, 6.8 diamonds) and for two values of F (shallow (a) and deep (b)). The curves with different colors are described in the text. (absolute n's values are given)*

It can be seen that the minimum value of the normalized integral E is clearly achieved in correspondance to different n's for each aspect ratio and also for each F values.

In particular, two values of F are analyzed, one shallow and a sligthly deeper F case. For each aspect ratio three curves are plotted. The different colors correspond to a different choice for the free parameters of the two relaxation region. The black curve corresponds to the choice : ($r_c$, $\alpha_1$, $\alpha_2$) = (0.39, 3, 2.1) (as in Fig.8), the red curve has set $\alpha_1$ equal to the on axis parallel current deduced by solving Eq.(3) (i.e. the single helical relaxation model is determining the value of the on axis parallel current density), the other two parameters are equal to the black curve case, finally the light blue curve has $\alpha_1$ as for the red case, while $\alpha_2$ is set to (0.6 $\alpha_1$) and $r_c$ is again left fixed (0.39). For the light blue case therefore the important degrees of freedom of the two relaxation model are, in practice, just reduced to one i.e. to the choice of the radius, $r_c$, which identify the position of the current sheets or derivative jumps. In Fig.10 the sensitivity to this parameter is analyzed for the shallow F case in the range $0.3 < r_c < 0.6$. It can be seen that only for $r_c$ =0.6 (i.e. a quite external reconnection layer) and for the highest R/a values, 5 and 6.8, a slight shift of the minimum is obseved toward n=8 (7) and 11 (10) respectively.

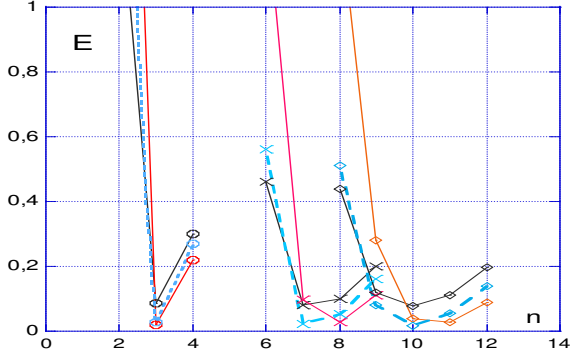

***Fig.10:*** *The integral difference E vs. n for different aspect ratios (R/a=2 circles, 5 crosses, 6.8 diamonds) for two values of $r_c$ (0.3 black and 0.6 red curve). A ligth blue dashed curve with $r_c$ =0.39 (as in Fig.9) is also plotted. (absolute n's values are given)*

.

It can therefore be concluded that the n values corresponding to the minima are not much sensitive to the free parameters of the two relaxation model and the behavior with both, aspect ratio and F, is in good agreement with the experimental results discussed in paper I (see also references therein).

Also the "fluctuations" of the predicted dominant modes (Fig.9(b) red curve at R/a=5 and black curve at R/a=6.8 and those in Fig.10 ) are compatible with experimental data where a certain variability of the dominant modes is observed especially a low F values.

The results displayed in Fig.9 and 10 appear to be quite robust. They have been obtained for parallel current profiles which remain finite at the plasma edge. Within the single helical relaxation theory this has been achieved by setting d=2 in Eq.(3) (and also by the appropriate setting of another free constant in the model), while, for the two region model, this is a quite natural outcome for a reasonable (i.e. experimentally relevant) choice of the free parameters. Importantly, as discussed in Sec.I for the case with d=2 all the relaxed states are stable and the F-Θ experimental operational region is well reproduced.

A further way of showing that by setting d=2 a good comparison with the experiments is obtained, is by looking at the predictions for the on axis resonant mode. Considering the shallow F case, in Fig.11 the inverse of the on axis q (corresponding, assuming, as done here, modes with m=1, to the theoretically predicted on axis resonant toroidal mode number) is plotted against the experimental dominant n mode for d=1 (as in Ref.[1]) and for d=2. It can be seen that the case with d=2 shows an almost perfect correspondance between the SH dominant n's and the on axis resonant mode predicted by the relaxation theory, while d=1 tends to underestimate the q on axis.

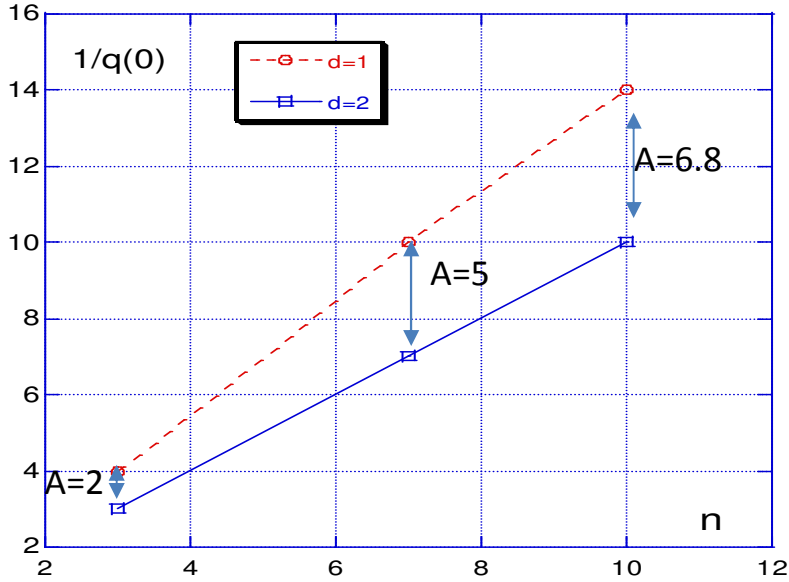

***Fig.11:*** *Relaxed states predicted on-axis modes for d=1 (as in Ref.[1], dashed line ) and for d=2 (plain line) vs. the experimental n values (the marks correspond to 3 aspect ratios, A= R/a= 2 , 5, 6.8 , at shallow F).*

To obtain the results diplayed in Fig.11 the experimental n values ( and aspect ratios) were used to calculate the relaxation model free parameter, $q_s$ (see Eq.(2) in section I). The figure shows that the choice of the free parameter d is quite important, in fact only for d=2 the predicted on axis resonant mode almost coincides with the experimental one (x and y values in the plot are equal), and also it suggests that it is not strictly necessary, at least in certain cases, to use the procedure of minimum deviation from the two region relaxed states, as done in Fig.9 and 10, to determine the relaxed states that predict the correct value of the first resonant mode (at least once that the experimental data are available). However, it should be emphasized that the procedure leading to the result displayed in Fig.9 and 10 is much more general since it includes the possibility of the existance in the plasma of reconnecting layers with current discontinuities and jumps, as discussed above Tearing stability, together with pressure and transport effects in the core plasma, could play a central role in the oscillatory time behavior of the SH states, as also recently discussed [12]. Within our two regions relaxation model, it would be very easy to introduce an assigned local jump of the parallel current profile that could, in turn, affect the global plasma stability [11] without the need of a whole rearrangement of the magnetic field profiles, as also strongly supported by the available experimental data. A mechanism producing a local effect such as an intensification of the current at (or near to) the resonance surface of the dominant mode, is just needed. Local heating can easily produce this phenomenon when assuming Spitzer's like plasma resistivity and Ohm's law. For example a chaos healing mechanism within the island [13] could be a possible candidate to explain a transport reduction and therefore the heating. The finite amount of the parallel current near to the wall, is likely instead, not so fundamental , and it is mainly a consequence of the attempt of reasonably matching the profiles obtained with the two region relaxation model. The fact that the role of the edge profiles is not a fundamental one, is shown on one hand from the relatively low sensitivity of the curves in Fig.9 from different choices of $\alpha_2$ ; on the other hand it has been verified that the minima given in Fig.9 and 10 are well reproduced even if the integral defining E is restricted over half of the plasma radius, i.e. it is calculated only in the core region excluding the edge plasma. This clearly indicates the importance of the profile rearrangements (and therefore of the physical effects that produce them) occurring in the core, in order to correctly predict the properties of the RFP's during the SH relaxation phases.

**Discusssion and Conclusions**

In a previous paper (paper I) it was found that a toroidal conductive shell has a preferential electromagnetic response to modes with toroidal mode numbers similar to those detected in experiments at various aspect ratios. However the sign of the corresponding helicity remained undetermined. The observation that helicities similar to those measured in experiments could "feel" a near ideal wall response, triggers the attention on the role of relaxation processes, as possible candidates to explain the behavior of the dominant helicities, including the determination of the correct sign of the dominant mode, both with the aspect ratio and with the reversal parameter, F .

Starting from this observation, an earlier proposed relaxation theory that assumes a dominant helicity in the plasma [1] has been reconsidered. Although interesting characteristics of the associated relaxed magnetic fields have emerged, whitin this theory there is no way of predicting the helicity that is favored in different conditions.

On the other hand, several reproducible and robust experimental observations in many machines strongly require a theory that is able to predict the dominant modes. We remind here some of the most relevant experimental references (see also paper I ). At aspect ratio (A=R/a) A=2 (Relax experiment) n=-3 and -4 dominant modes are detected [14]. At A= 2.88, MST (Madison Symmetric Torus) finds dominant n=-5 or -6 modes [15,16]. The RFX-mod device at A=4.3 finds mostly dominant n=-7 [17,18] . At A= 6.88 the Extrap-T2R machine detects n=-11,-12,-13 depending on the F parameter [19,20].

The presence in the plasma of a reconnecting mode with an associated island structure (typical of the SH states) suggests the necessity to develop a two region relaxation model that generalizes Taylor's states. In this model, in addition, current and field discontinuities can be easily included, a feature that has been shown to directly affects the system stability [11]. Then, it has been hypothesized that the observed helicity in experiments could be the one that minimize the "distance" between the magnetic profiles obtained from the single helical relaxation, on one hand, and from the two region relaxation theory on the other. The assumption of minimum distance is based on the observation that current jumps are characteristics that emerge in experiments in an unpredictable and discontinous way. Therefore the "smooth" relaxed states obtained from [1,9], under this point of view, could be considered as a sort of "average" configurations, among a series of more general states, where the presence of island-like structures could play an important role. As shown in this work, this hypothesis, is able to predict in a simple way the behavior of the dominant helical mode as a function of aspect ratio and F (reversal parameter) values.

Our results seem to indicate that in the core of the RFP's plasma a full relaxation, corresponding to a flat Taylor's like parallel current density profile, is present. However, in the region between the wall (r/a=1) and the dominant reconnecting mode radius (assuming that one dominant mode exists), r= $r_c$, only a partial relaxation is achieved and there the current density profile is varying with the radius.

As shown, the overall average parallel current density profile, from the core to the edge, is well described by the single helical relaxation model [1,9]. However to predict the dominant modes at different aspect ratios and reversal parameters it has been shown that the Taylor's like relaxation should also be taken into account, in the core region, $0 < r < r_c$.

Even if it was not possible in this work to justify rigourously the fact that the dominant n's correspond to the minimum distance (minimum E value) between the two models, a possible simple physical interpretation, also coherent with the observations, is that the presence of a dominant reconnecting mode could drive a full reconnection in the RFP core which is clearly very well described by the two region relaxation model proposed here. In this case the two models presented in the paper , i.e the single helical relaxation and the two region relaxation, should ' a fortiori' predict the same final states in order to agree with the experiments.

It has been shown that this is obtained only for specific n's values under different geometrical (aspect ratio) or physical (F parameter) conditions. Importantly each of the two models, considered separatlely, are not able to make full predictions.

The choice of the free parameter d=2 (see Eq.(1) and (3)) (instead of d=1 as in [1,8,9]) has also revelead to be quite important in achieving this result, since it predicts almost correctly the value of the on-axis q, and therefore of the first resonant mode (at least at shallow F).

By looking at references [14-20] a dominant mode with n=-3,- 7 and -11 is observed in experiments (at shallow F) with R/a=2, 4.3 and 6.8 respectively, while these numbers shift to n=-4, -8 , -12 (or -13) at a deeper reversal, in good agrrement with the present results (see Fig.9).

These helicities correspond to modes resonant near the magnetic axis (at r=0), the so called core resonant modes. Their emergence is strictly related to the fact that the parallel current profile has a finite gradient in the central region, which is allowed both by the single helical and by the two regions relaxation models discussed here.

Although the two region relaxation model allows the presence of current discontinuity, we also verified that the results are not influenced for small or moderate current jump amplitudes, therefore in the paper we only consider the contimous limit. Current discontinuities and also current sheets within the islands (not treated here) could however determine the system stability and the relaxation oscillations observed in experiments, as for example described in the model developed in Ref. [12].

To conclude and summarize: In this paper it has been shown that 2 types of relaxations could play a role in the emergenge of the SH states in RFPs: on one side, a smooth relaxation which preserves some global invariants in systems where an ideal or quasi ideal boundary is present, and, on the other hand, a relaxation process where the presence of magnetic islands ( justifying a multiple region rleaxation process) and local effects, are important. According to our analysis the states that are actually achieved are those for which there is a minimal deviation between the magnetic fields (and associated currents) predicted by these two models. We found in particular that for each aspect ratio and degree of toroidal field reversal this minimum is achieved in correspondance to specific mode numbers in very good agreement with experimental measurements in different RFP devices. The relaxation models studied here correspond to partially relaxed states with a non flat parallel current density and are able to match satisfactorily the operational regions of the existing RFP machines. The development of current sheets or current jumps can, in turn, affect the global stability of the system and may explain, when associated with a suitable reconnection and transport models, the sawtoothing behavior seen in experiments [12].

The theory developed in this paper can account, for the first time, for the observed dominant mode numbers in experiments in different conditions, while the results of paper I explain why for these modes the ideal wall boundary condition is well satisfied. Note also that the results of paper I are obtained in toroidal geometry. The two papers can give therefore a new and more general point of view, for example with respect to numerical simulations, regarding the interpretation of the experimental observations of the SH states in RFP's . In fact, it should be considered that fine structures like current sheets and/or the effect of the toroidal geometry for the shell response are difficult aspects to be handled within global numerical simulations, and therefore semi-analytical theories and alternative approaches, as that outlined here, are certainly useful and can furnish very helpful physical interpretation guidelines.

**Acknowledgments:** I am grateful to my colleague Italo Predebon for his advice in installing under Linux some python libraries useful to solve eigenvalue problems. I thanks also P. Zanca for helpful comments.